\documentclass{mem}
\usepackage{natbib}
\usepackage[varg]{txfonts}
\usepackage{graphicx}
\usepackage[a4paper]{hyperref}
\idline{75}{282}
\begin{document}
\def\teff{$T_{\rm eff}$}
\def\logg {log\,g}
\def\lambo{$\lambda$ Boo }
\def\vsini {$v\,\sin i$ }
\def\kms {$\mathrm{km\, s^{-1}}$ }
\newcommand{\pun}[1]{\,#1}
\newcommand{\loggf}{\ensuremath{\log\,gf}}
\newcommand{\mlp}{\ensuremath{\alpha_{\mathrm{MLT}}}}
\newcommand{\LHD}{{\sf LHD}}
\newcommand{\xx}{\ensuremath{\mathrm{1D}_{\mathrm{LHD}}}}
\newcommand{\cobold}{{\sf CO$^5$BOLD}}
\newcommand{\mD}{\ensuremath{\left\langle\mathrm{3D}\right\rangle}}
\newcommand{\linfor}{{\sf Linfor3D}}

\title{
Micro- and macroturbulence derived from 3D~hydrodynamical 
stellar atmospheres
}

   \subtitle{}

\author{
M. Steffen\inst{1}
\and
H.-G.~Ludwig\inst{2,3}
\and E.~Caffau\inst{3}
       }

\offprints{msteffen@aip.de}

\institute{ 
Astrophysikalisches Institut Potsdam, An der Sternwarte 16, D-14482 Potsdam, 
Germany 
\and
CIFIST Marie Curie Excellence Team
\and
GEPI -- Observatoire de Paris, CNRS, Universit\'e Paris Diderot; 92195
Meudon, France
}

\authorrunning{Steffen et al. }

\titlerunning{Micro- and macroturbulence}

\abstract{
The theoretical prediction of micro- and macroturbulence ($\xi_{\rm mic}$ 
and $\xi_{\rm mac}$) as a function of stellar parameters can be useful for 
spectroscopic work based on 1D model atmospheres in cases where an empirical
determination of $\xi_{\rm mic}$ is impossible due to a lack of suitable 
lines and/or macroturbulence and rotational line broadening are difficult to
separate. In an effort to exploit the CIFIST 3D model atmosphere grid for 
deriving the theoretical dependence of $\xi_{\rm mic}$ and $\xi_{\rm mac}$ 
on effective temperature, gravity, and metallicity, we discuss different 
methods to derive $\xi_{\rm mic}$ from the numerical simulations, and report 
first results for the Sun and Procyon. In both cases the preliminary analysis 
indicates that the microturbulence found in the simulations is significantly 
lower than in the real stellar atmospheres.

}
\maketitle{}

\section{Introduction}
3D hydrodynamical simulations of stellar surface convection provide a
physically self-consistent description of the non-thermal velocity
field generated by convection, overshoot, and waves. This is one of
the great advantages over classical 1D model atmospheres where the
properties of the photospheric velocity field need to be specified
empirically in terms of the free parameters $\xi_{\rm mic}$ and 
$\xi_{\rm mac}$. 
Even if these parameters are irrelevant in the context
of 3D hydrodynamical model atmospheres, the latter may be used to
predict the magnitude of $\xi_{\rm mic}$ and $\xi_{\rm mac}$ as
a function of the stellar parameters (\teff, \logg, [M/H]).
The CIFIST 3D model atmosphere grid (Ludwig et al., this volume)
provides a suitable database for this purpose.
In the following, we investigate different methods to derive the 
parameter $\xi_{\rm mic}$ (and subsequently $\xi_{\rm mac}$) from 
our 3D model atmospheres which were computed with the \cobold\ 
code\footnote{http://www.astro.uu.se/$\sim$bf/co5bold\_main.html} 
\citep{Freytag+al02,Wedemeyer+al04}. The Sun (\teff=5780~K, \logg=4.44,
[M/H]=0) and Procyon (\teff=6500~K, \logg=4.0, [M/H]=0) serve as 
benchmarks for the present study.

\section{Methods to derive $\xi_{\rm mic}$ and $\xi_{\rm mac}$}
All methods described below presume that the photospheric 
velocity field may be characterized by a single, depth-independent value 
of $\xi_{\rm mic}$ and $\xi_{\rm mac}$, and that both micro- and 
macroturbulence have an isotropic Gaussian probability distribution of 
the line-of-sight velocity, $P(v) \sim \exp(-v^2/\xi^2)$. Synthetic 
spectral lines serve as a diagnostic tool to probe $\xi_{\rm mic}$ and 
\,$\xi_{\rm mac}$ via the total absorption (equivalent width $W$) and the
shape of the line, respectively.

{\bf Method~1} \emph{(M1)} 
is considered the most accurate procedure to extract the 
microturbulence parameter from a 3D numerical convection simulation. 
Unlike the other methods described below, it relies only on the 3D model,
and yields a value of $\xi_{\rm  mic}$ for any individual spectral line, 
thus allowing to map $\xi_{\rm  mic}(W)$. Given the spectral line 
parameters, the line profile is computed from the 3D model with different 
velocity fields: (i) using the original 3D hydrodynamical velocity field, 
and (ii) replacing the 3D hydrodynamical velocity field by an isotropic,
depth-independent microturbulence, like in classical 1D spectrum
synthesis, but retaining the full 3D thermodynamic structure. Now the
microturbulence associated with the considered spectral line,
$\xi^{(1)}_{\rm mic}$, is defined by the requirement of matching line
strengths: $W_{\rm 3D}({\rm hydro}) = W_{\rm 3D}(\,\xi^{(1)}_{\rm mic})$, 
where $W_{\rm 3D}({\rm hydro})$ and $W_{\rm 3D}(\,\xi_{\rm mic})$ 
are the equivalent widths obtained in steps (i) and (ii), respectively.

Once $\xi_{\rm mic}$ is determined as explained above, the
macroturbulence associated with the considered spectral line is
defined by minimizing the mean square difference between the 3D line
profile computed in step (i) with the hydrodynamical velocity field,
and the modified 3D line profile obtained in step (ii) with
$\xi^{(1)}_{\rm mic}$ and subsequent convolution with the macroturbulence
velocity dispersion ($\xi_{\rm mac}$). The IDL procedure MPFIT 
is used to find the solution $\xi^{(1)}_{\rm mac}$ that gives the best
fit to the original 3D profile.

{\bf Method~2a/b} \emph{(M2a/b)} 
The idea of this method is to replace the modified 3D models
(furnished with a classical microturbulence velocity field) used in
step (ii) of \emph{Method~1} with an arbitrary 1D model atmosphere. 
However, this concept does not work for a single spectral line, but
instead has to rely on a set of spectral lines ranging from weak to partly
saturated. In \emph{Method~2} the set of lines is generated from a
\emph{curve-of-growth}, i.e. all lines share the same atomic parameters
except for the $f$-value, which controls the line strength.
Given the set of \emph{fictitious} spectral lines and a 1D model 
atmosphere, we first compute for each line $i$ the 1D-3D abundance difference 
$\Delta \log \epsilon_i(\,\xi_{\rm mic})$,
defined by the condition $W_{\rm 3D}\,(\log g f_i) = 
W_{\rm 1D}\,(\xi_{\rm mic}, \log g f_i + \Delta \log \epsilon_i)$,
i.e. $\Delta \log \epsilon_i$ is the logarithmic abundance difference between
the abundance derived from the 1D model by fitting the equivalent of the
3D profile, and the true abundance assumed in computing the 3D spectral line.
In \emph{M2a} we compute the slope of the linear regression to the data set
$\{W_{\rm 3D},\; \Delta \log \epsilon_i(\,\xi_{\rm mic})\}$,
and define $\xi^{(2a)}_{\rm mic}$ by the condition of vanishing slope,
$\beta(\,\xi^{(2a)}_{\rm mic})=0$. Alternatively, in \emph{M2b} the 
microturbulence is required to produce the minimum standard deviation of 
\{$\Delta \log \epsilon_i$\}, $\sigma(\,\xi^{(2b)}_{\rm mic})$\,=\,$\min$.
Once $\xi_{\rm mic}$ is determined, the macroturbulence associated with 
each individual spectral line, $\xi^{(2)}_{\rm mac}$, is found by fitting
the original 3D line profile with the corresponding 1D line profile 
(with fixed $\xi^{(2)}_{\rm mic}$) through variation of $\xi_{\rm mac}$, 
in essentially the same way as in \emph{Method~1}.

{\bf Method~3a/b} \emph{(M3a/b)}
is very similar to \emph{Method~2}, except for utilizing
a sample of \emph{real} spectral lines of different strength 
(and different wavelength, excitation potential, etc.) instead of a 
set of fictitious lines lying on a single curve-of-growth.
Adjusting the value of $\xi_{\rm mic}$ to minimize the difference 
in $\Delta \log \epsilon_i$
between weak and strong lines obtained from the preferred 1D model results 
in $\xi^{(3a)}_{\rm mic}$ (zero slope condition). Similarly, 
minimizing the overall dispersion of $\Delta \log \epsilon_i$ gives
$\xi^{(3b)}_{\rm mic}$. \emph{Method~3} corresponds to the classical 
definition of $\xi_{\rm mic}$. Again $\xi^{(3)}_{\rm mac}$ is found as 
in \emph{Method~2}.

Note that all 3 methods have the advantage that errors in $\log g f$
cancel out. Only \emph{Method~3} can also be applied to observed stellar
spectra; in this case the accuracy of the $\log g f$ values \emph{is} 
crucial. Obviously, the results depend on the selected spectral lines, 
and, for \emph{Methods~2} and \emph{3}, on the choice of the 1D model 
atmosphere.

In principle, it should be possible to derive $\xi_{\rm mic}$ and
$\xi_{\rm mac}$ directly from evaluating the 3D hydrodynamical
velocity field without resorting to synthetic spectral lines. 
But a suitable procedure ({\bf Method~4}) has yet to be developed.

\section{First results: Sun and Procyon}

\begin{figure*}[]
\mbox{\includegraphics[bb=45 32 580 390, clip=true, width=\hsize]
{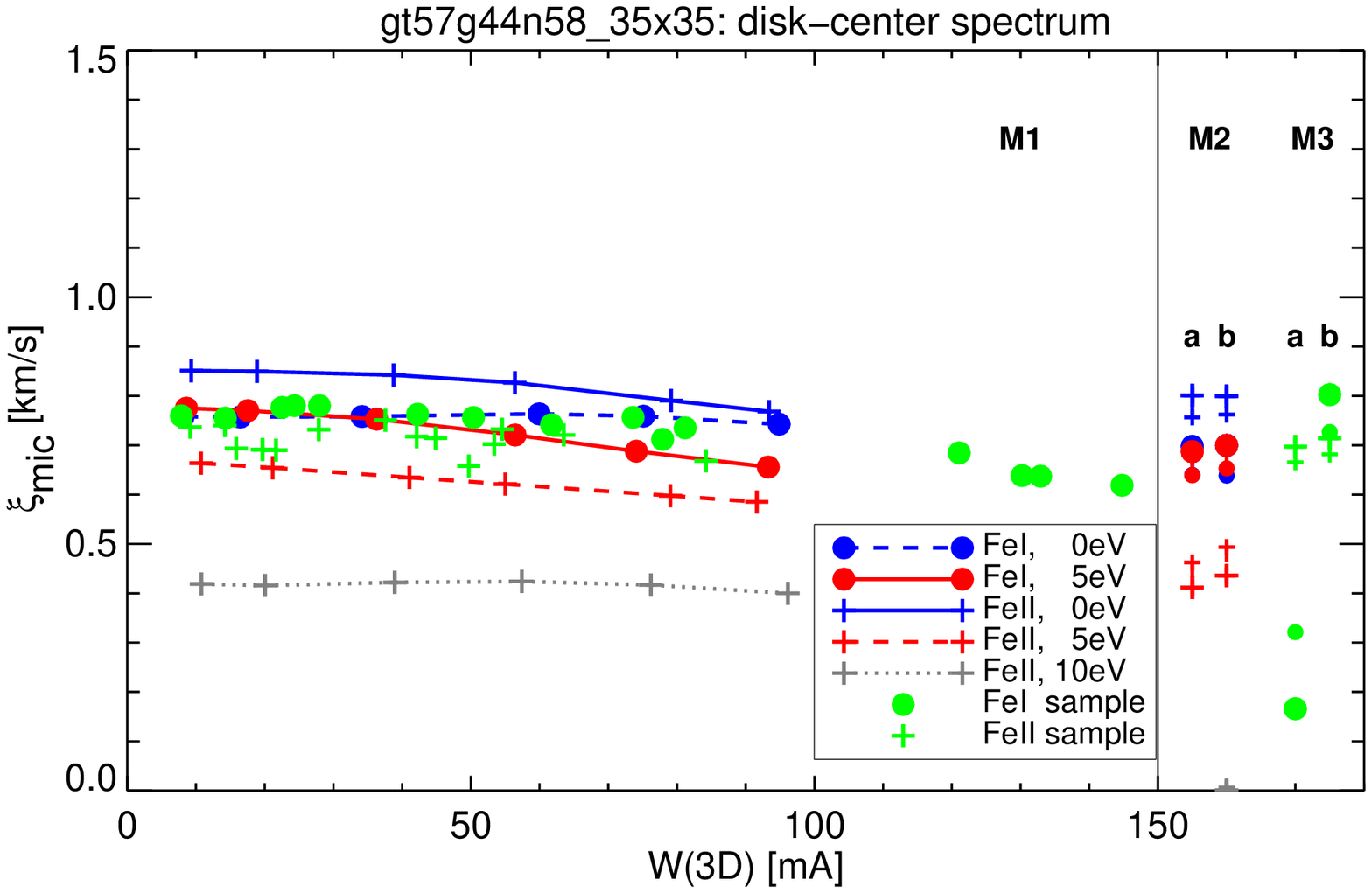}}
\mbox{\includegraphics[bb=45 32 580 390, clip=true, width=\hsize]
{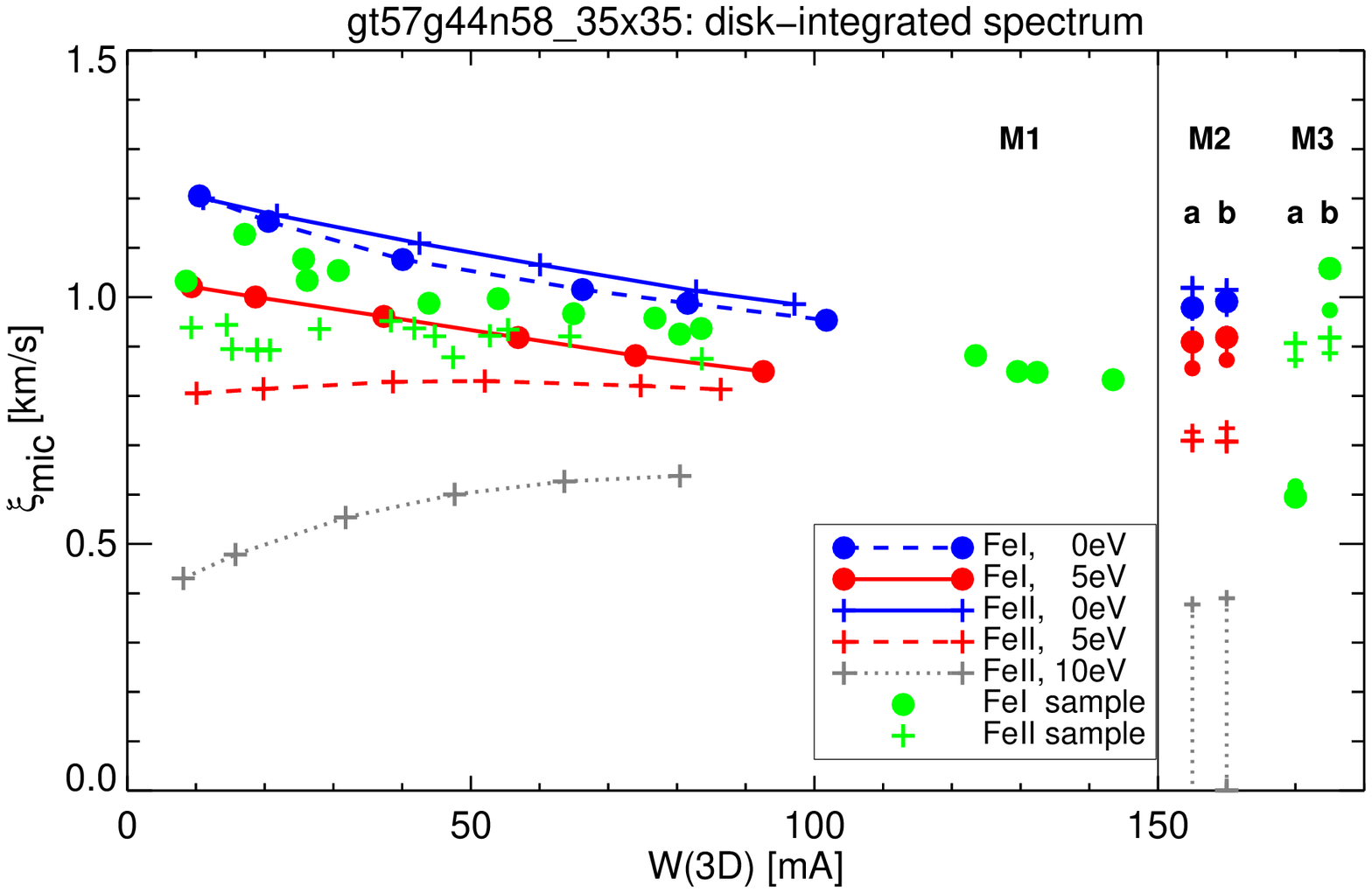}}
\caption{\footnotesize
Overview of $\xi_{\rm mic}$ derived from our standard 3D solar model
atmosphere (gt57g44n58) determined with \emph{M1, M2,} and \emph{M3}
for a total of 60 different Fe I (filled circles) and Fe II (plus signs) lines.
Connected symbols represent fictitious lines sharing the same curve-of-growth, 
individual symbols indicate real spectral lines. The results of \emph{M2a,b} 
(red, blue) and \emph{M3a,b} (green) are shown to the right of the vertical 
line in each panel for two different 1D models (smaller symbols: 
$\langle$3D$\rangle$, larger symbols: HM). Top and bottom panels refer to 
disk-center and full-disk spectra, respectively.
}
\label{fig1}
\end{figure*}

\begin{figure*}[]
\mbox{\includegraphics[bb=45 32 580 390, clip=true, width=\hsize]
{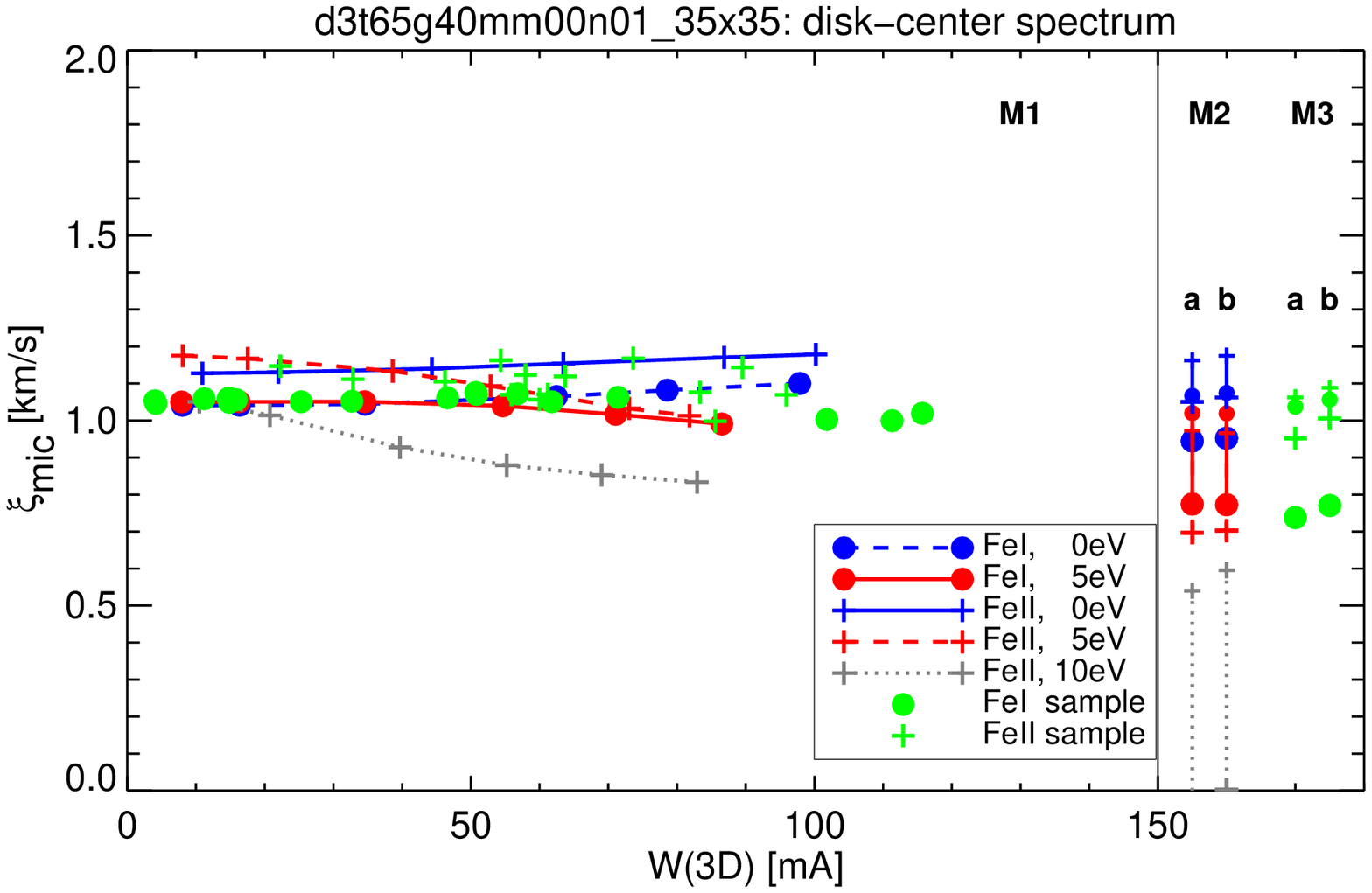}}
\mbox{\includegraphics[bb=45 32 580 390, clip=true, width=\hsize]
{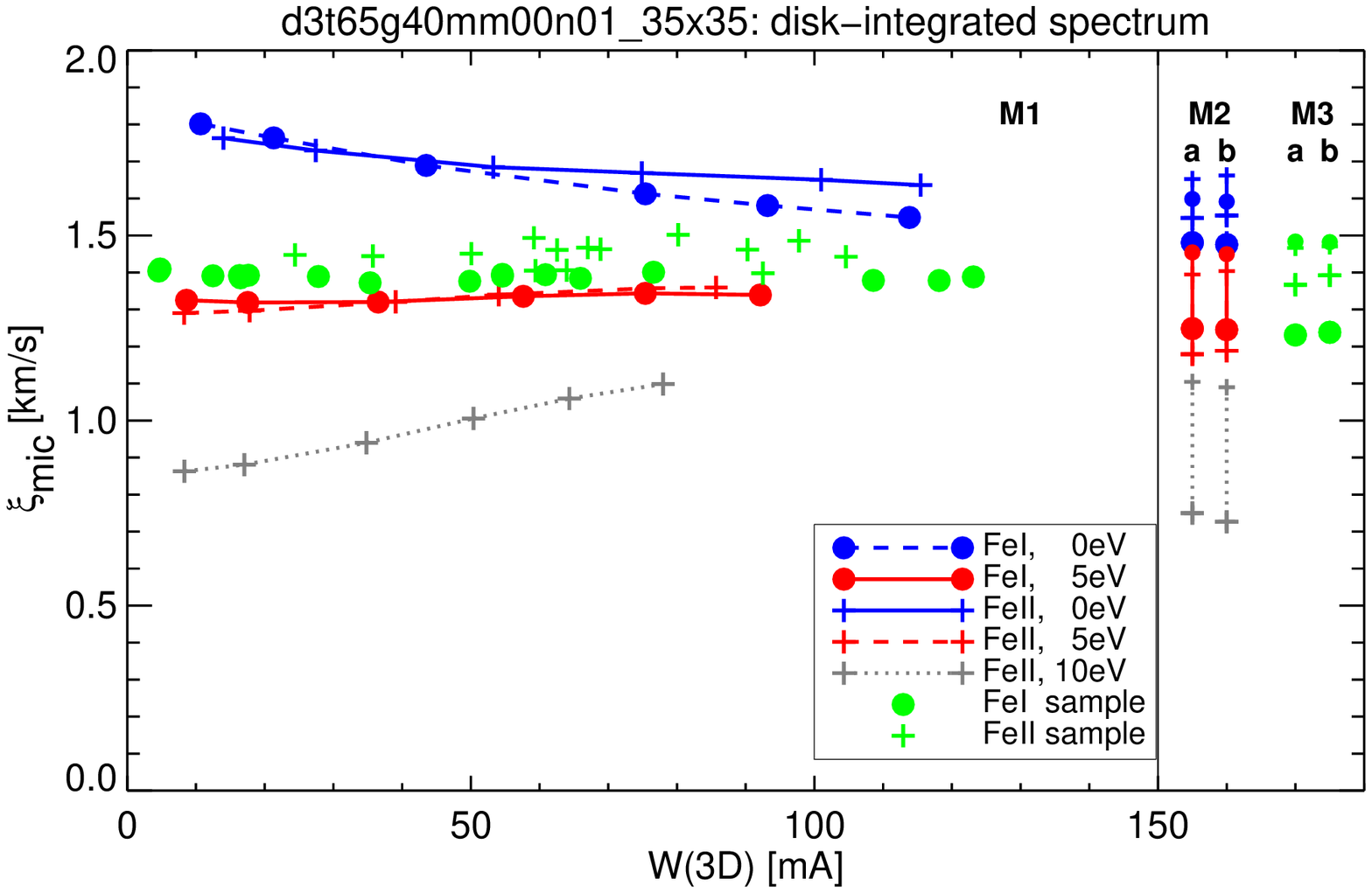}}
\caption{\footnotesize
As Fig.\,\ref{fig1}, but for a 3D model atmosphere representative of Procyon, 
using a total of 56 different Fe I (filled circles) and Fe II (plus signs) 
lines. 1D models used with \emph{M2, M3}: $\langle$3D$\rangle$ (smaller 
symbols) and LHD [$\alpha_{\rm MLT}$=$0.5$] (larger symbols). Top and bottom 
panels refer to disk-center and full-disk spectra, respectively.
}
\label{fig2}
\end{figure*}

Figures \ref{fig1} and \ref{fig2} show the determination of the 
microturbulence parameter from a 3D model atmosphere of the
Sun and Procyon, respectively, according to the three methods described 
above. We considered 5 sets of fictitious iron lines (Fe~I 0, and 5~eV, Fe~II
0, 5, and 10~eV, used with \emph{M1, M2}), and one sample of real Fe~I and 
Fe~II lines each (used with \emph{M1, M3}). The most obvious result is that
the microturbulence derived from the flux spectra is systematically higher
than that obtained from the intensity spectra, in agreement with observational
evidence \citep[e.g.][]{Holweger+al78}. Moreover, the derived value of 
$\xi_{\rm mic}$ depends on the type and strength of the considered spectral
line. \emph{Method~1} clearly reveals that high-excitation lines `feel' a
lower microturbulence than low-excitation lines. The dependence of 
$\xi_{\rm mic}$ on line strength is non-trivial (see Fig.\,\ref{fig1}, 
lower panel).
In general, \emph{Methods~2a} and \emph{2b} give very similar results,
which moreover agree well with the results obtained from \emph{Method~1} for
the stronger lines on the curve-of-growth. In contrast, \emph{Methods~3a} 
and \emph{3b} can give rather discordant answers, if the sample of spectral
lines is not sufficiently homogeneous (see again lower panel of
Fig.\,\ref{fig1}).

\begin{figure}[]
\mbox{\includegraphics[bb=54 360 558 720, clip=true, width=\hsize]
{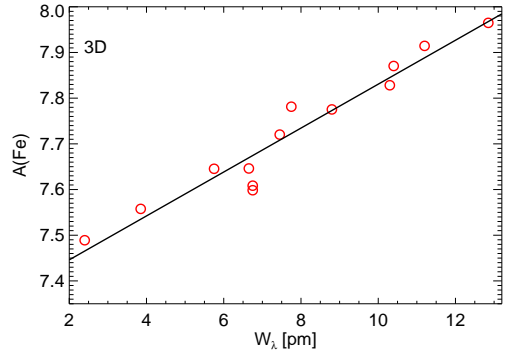}}
\caption{\footnotesize
Iron abundance derived from individual Fe~II lines using a 3D \cobold\ model 
with \teff=6500~K, log~g=4.0, [M/H]=0, with log $gf$ values taken from 
\citet{mb09}, and van der Waals line broadening treated according to 
\citet{abo4}. The strong trend of abundance with equivalent width indicates 
an apparent lack of small-scale turbulence in the hydrodynamical velocity 
field of the Procyon model.
}
\label{fig3}
\end{figure}

In Table\,\ref{tab1}, the theoretical predictions of micro- and
macroturbulence based on our hydrodynamical model atmospheres for the
Sun and Procyon are summarized and confronted with empirical results
from the literature. This preliminary analysis indicates that the
theoretical predictions of $\xi_{\rm mic}$ fall significantly below
the classical empirical estimates, for both solar intensity and flux
spectra, and even more clearly for Procyon, as demonstrated
unmistakably in Fig.\,\ref{fig3}.  On the other hand, the
macroturbulence derived from the \cobold\ models is somewhat larger
than deduced from observations, such that the \emph{total} non-thermal
rms velocity $v_{\rm turb}$ (columns (6) and (7) of Tab.\,\ref{tab1}),
and hence the total line broadening, is very similar in simulations
and observations.

\begin{table*}[]
\caption{Empirical values of $\xi_{\rm mic}$ and $\xi_{\rm mac}$ from the 
literature, compared with the theoretical results derived in this work from 
3D \cobold\ hydrodynamical model atmospheres for the Sun and Procyon (weighted 
average over \emph{Methods~1, 2} and \emph{3}).}
\label{tab1}
\fontsize{8}{10}\selectfont
\begin{center}
\begin{tabular}{lcccccc}
\noalign{\smallskip}\hline\noalign{\smallskip}
  Atmosphere / & \multicolumn{2}{c}{$\xi_{\rm mic}$ [km/s]} & 
                 \multicolumn{2}{c}{$\xi_{\rm mac}$ [km/s]} &
                 \multicolumn{2}{c}{$v_{\rm turb}=
                 \sqrt{(\xi_{\rm mic}^2+\xi_{\rm mac}^2)/2}$~~[km/s]}\\
  Model        & disk-center      & full-disk   & 
                 disk-center      & full-disk   &
                 disk-center      & full-disk   \\
\noalign{\smallskip}\hline\noalign{\smallskip}
Sun, observed$^{\,a}$     & $1.00\pm 0.15$  & $1.35\pm 0.15$  
                          & $1.63\pm 0.15$  & $1.90\pm 0.15$ 
                          & $1.35\pm 0.10$  & $1.65\pm 0.10$  \\
3D solar model            & $0.70\pm 0.10$  & $0.95\pm 0.15$  
                          & $1.85\pm 0.45$  & $2.30\pm 0.30$ 
                          & $1.40\pm 0.30$  & $1.75\pm 0.20$  \\
(gt57g44n58)              &           &          &        &   \\
\noalign{\smallskip}\hline\noalign{\smallskip}
Procyon, observed$^{\,b}$ &  ---            & $2.10\pm 0.30$  
                          &  ---            & $4.20\pm 0.50$ 
                          &  ---            & $3.30\pm 0.30$  \\
3D Procyon model          & $0.95\pm 0.25$  & $1.45\pm 0.25$ 
                          & $3.10\pm 0.40$  & $4.45\pm 0.20$ 
                          & $2.30\pm 0.30$  & $3.30\pm 0.15$  \\
(d3t65g40mm00n01)         &           &          &        &   \\
\noalign{\smallskip}\hline\noalign{\smallskip}
\end{tabular}
\end{center}
Notes: a: \citet{Holweger+al78}, full-disk values interpolated: 
$\xi^2$(full-disk)=$\xi^2$(disk-center)/2 + $\xi^2$(limb)/2;\\ 
b: \citet{Steffen85}
\end{table*}

\section{Conclusions}
We have developed different methods to extract the parameters $\xi_{\rm mic}$
and $\xi_{\rm mac}$ from 3D hydrodynamical simulations, and found all methods
to give consistent results. As expected, $\xi_{\rm mic}$ and $\xi_{\rm mac}$
depend systematically on the properties of the selected spectral lines.
Our preliminary analysis based on state-of-the-art \cobold\ models for the
Sun and Procyon indicates that the microturbulence seen in the simulations 
is significantly lower than measured in the real stars. This suggests
that the velocity field provided by the current 3D hydrodynamical
models is less `turbulent' than it is in reality. It seems that the
simulations predict essentially the correct total rms velocity, while
underestimating the small-scale and overestimating the large-scale 
fluctuations. If confirmed, this deficiency implies a systematic overestimation
of 3D abundances from stronger lines. Apparently, our findings are in
conflict with the conclusions by \citet{A2000}. Further investigations are 
necessary to find the basic cause of this disagreement and to clarify possible
implications for high-precision abundance studies.


\bibliographystyle{aa}

\end{document}